\documentclass{emulateapj}
\usepackage{lscape}
\usepackage{subfigure}
\usepackage{graphicx}
\usepackage{epstopdf}
\usepackage{longtable}
\usepackage{amsmath}
\usepackage{amssymb}
\usepackage{bigints}

\newcommand{\noprint}[1]{}

\newcommand{\ik}{{\it Kepler~}}

\newcommand{\koi}{{KOI-256~}}

\newcommand{\vsini}{{$V \sin i$~}}
\newcommand{\teff}{$T_{\rm eff}$~}
\newcommand{\kms}{{km~$\rm s^{-1}$}}

\def\caii{Ca\,II~}

\begin{document}

\title{Characterizing the Cool KOI\lowercase{s}. V.  KOI-256: A Mutually Eclipsing Post-Common Envelope Binary}

\author{Philip S. Muirhead\altaffilmark{1},
 Andrew Vanderburg\altaffilmark{2,11},
  Avi Shporer \altaffilmark{1}, 
  Juliette Becker\altaffilmark{1,11,12}, 
  Jonathan J. Swift\altaffilmark{1}, 
  James P. Lloyd\altaffilmark{3}, 
  Jim Fuller\altaffilmark{3}, 
  Ming Zhao\altaffilmark{4}, 
  Sasha Hinkley\altaffilmark{1,13}, 
  J. Sebastian Pineda\altaffilmark{1},
   Michael Bottom\altaffilmark{1}, 
   Andrew W. Howard\altaffilmark{5}, 
   Kaspar von Braun\altaffilmark{6,7},
    Tabetha S. Boyajian\altaffilmark{8},
     Nicholas Law\altaffilmark{9}, 
     Christoph Baranec\altaffilmark{1}, 
     Reed Riddle\altaffilmark{1}, 
     A. N. Ramaprakash\altaffilmark{10}, 
     Shriharsh P. Tendulkar\altaffilmark{1}, 
     Khanh Bui\altaffilmark{1}, 
     Mahesh Burse\altaffilmark{10}, 
     Pravin Chordia\altaffilmark{10}, 
     Hillol Das\altaffilmark{10}, 
     Richard Dekany\altaffilmark{1}, 
     Sujit Punnadi\altaffilmark{10} and 
     John Asher Johnson\altaffilmark{1,6}}

\altaffiltext{1}{California Institute of Technology, 1200 East California Boulevard, Pasadena, CA  91125, USA}
\altaffiltext{2}{University of California, Berkeley, CA 94720, USA}
\altaffiltext{3}{Cornell University, Ithaca, NY 14850, USA}
\altaffiltext{4}{Department of Astronomy and Astrophysics, 525 Davey Laboratory, The Pennsylvania State University, University Park, PA 16802, USA}
\altaffiltext{5}{Institute for Astronomy, University of Hawaii, 2680 Woodlawn Drive, Honolulu, HI 96822, USA}
\altaffiltext{6}{NASA Exoplanet Science Institute, California Institute of Technology, MC 100-22, Pasadena, CA 91125, USA}
\altaffiltext{7}{Max Planck Institute for Astronomy, K\"{o}nigstuhl 17, D-69117 Heidelberg, Germany}
\altaffiltext{8}{Department of Astronomy, Yale University, New Haven, CT 06511, USA}
\altaffiltext{9}{Dunlap Fellow, Dunlap Institute for Astronomy and Astrophysics, University of Toronto, 50 St. George Street, Toronto M5S 3H4, Ontario, Canada}
\altaffiltext{10}{Inter-University Centre for Astronomy \& Astrophysics, Ganeshkhind, Pune, 411007, India}
\altaffiltext{11}{Caltech Summer Undergraduate Research Fellow (SURF)}
\altaffiltext{12}{Alain Porter Memorial SURF Fellow}
\altaffiltext{13}{NSF Astronomy and Astrophysics Fellow}

\slugcomment{Accepted for publication in the Astrophysical Journal}

\begin{abstract}

We report that {\it Kepler} Object of Interest 256 (KOI-256) is a mutually eclipsing post-common envelope binary (ePCEB), consisting of a cool white dwarf ($M_\star = 0.592 \pm 0.089 M_\Sun$, $R_\star = 0.01345 \pm 0.00091 \, R_\Sun$, \teff = $7100 \pm 700 \, K$) and an active M3 dwarf ($M_\star = 0.51 \pm 0.16 M_\Sun$, $R_\star = 0.540 \pm 0.014 R_\Sun$, \teff = $3450 \pm 50\,K$) with an orbital period of 1.37865 $\pm$ 0.00001 days.  KOI-256 is listed as hosting a transiting planet-candidate by Borucki et al. and Batalha et al.; here we report that the planet-candidate transit signal is in fact the occultation of a white dwarf as it passes behind the M dwarf.  We combine publicly-available long- and short-cadence {\it Kepler} light curves with ground-based measurements to robustly determine the system parameters.  The occultation events are readily apparent in the {\it Kepler} light curve, as is spin-orbit synchronization of the M dwarf, and we detect the transit of the white dwarf in front of the M dwarf halfway between the occultation events.  The  size of the white dwarf with respect to the Einstein ring during transit ($R_{\rm Ein} = 0.00473 \pm 0.00055 \, R_\Sun$) causes the transit depth to be shallower than expected from pure geometry due to gravitational lensing.  KOI-256 is an old, long-period ePCEB and serves as a benchmark object for studying the evolution of binary star systems as well as white dwarfs themselves, thanks largely to the availability of near-continuous, ultra-precise {\it Kepler} photometry. 

\end{abstract}


\maketitle

\section{Introduction}

Post-common-envelope binaries (PCEBs) serve as important laboratories for studying the long-term evolution of binary star systems \citep[e.g.][]{Rebassa2007, Schreiber2010, Zorotovic2010, Zorotovic2011, Morgan2012}.  Consisting of a co-orbiting white dwarf and main-sequence star, PCEBs are presumed to have evolved from a stage when the main-sequence star orbited within the envelope of the white dwarf progenitor, an asymptotic giant branch (AGB) star \citep[e.g.][]{Paczynski1976, Willems2004}.  In this scenario, the main-sequence star loses angular momentum as it orbits within the envelope, eventually leaving a main-sequence star and white dwarf in a short-period orbit \citep[e.g.][]{Nordhaus2012, Spiegel2012}.  A PCEB will continue to lose angular momentum via magnetic braking and gravitational waves and will eventually begin mass-transfer, resulting in a cataclysmic variable.  For this reason, PCEBs are sometimes referred to as pre-cataclysmic variables \citep[pre-CVs, e.g.][]{Schreiber2003}.  To date, thousands of PCEBs have been discovered spectroscopically \citep[e.g.][]{Rebassa2010}.

{\it Eclipsing} PCEBs (ePCEBs) are much rarer, with several dozen detected to date \citep[][]{Drake2010,Law2012,Parsons2012}.  Such systems allow for precise measurements of the orbiting stars' physical parameters.  By combining precision photometric measurements of the white dwarf occultation events with radial velocity measurements, the masses, radii and temperatures of the component stars can be determined to relatively high accuracy and precision with few assumptions.  

In this paper, we show that {\it Kepler} Object of Interest 256 (KOI-256) \footnote{KIC 11548140, 2MASS J19004443+4933553, $\alpha$=285.$^{\circ}$18513, $\delta$=+49.$^{\circ}$56537} is an ePCEB consisting of an active M3 dwarf (dMe) and a cool white dwarf.  KOI-256 was listed as hosting a transiting planet candidate in \citet[][]{Borucki2011b} and \citet[][]{Batalha2013}, referred to as KOI-256.01, with an orbital period of 1.3786789 days and a fractional transit depth of 2.42 \%.  \citet{Batalha2013} note that the long-cadence planet-candidate transit light-curve appears to be V-shaped: an indication of a false-positive.  In the following sections, we show that the planet-candidate transit is in fact the occultation of a white dwarf as it travels behind an M dwarf, and that KOI-256.01 is therefore a false positive planet candidate.  This is confirmed by large sinusoidal radial velocity variations in the M dwarf spectrum synchronous with the occultation ephemeris, a trapezoidal occultation light curve with sharp ingress and egress, and a weak transit signal occurring halfway between the occultation events.  The depths of the occultation and transit are consistent with an eclipsing white dwarf, including a shallower transit depth than expected from geometry alone due to gravitation lensing.  With an orbital period of greater than 1 day, KOI-256 is a relatively long period ePCEB \citep[][]{Drake2010, Gomez2011, Law2012, Parsons2012}.  As we show in Section \ref{teff}, the white dwarf is also cool (\teff = 7100 $\pm$ 700 $K$), consistent with an old white dwarf.

In a previous letter, \citet{Muirhead2012b} characterized KOI-256 as an M3 dwarf with \teff = 3450 $\pm$ 50 $K$ and [M/H] = +0.31 $\pm$ 0.10 using the near-infrared spectroscopic indices of \citet{Rojas2012}.  They interpolated those values onto the Dartmouth evolutionary isochrones \citep[][]{Dotter2008,Feiden2011} to estimate the mass ($0.43 \pm 0.06 M_\Sun$) and radius ($R_\star = 0.42 \pm 0.06 R_\Sun$) of the M dwarf.  However, that analysis assumed KOI-256 was a field M dwarf, and the derived mass and radius did not consider the effects of PCEB evolution, rapid rotation or stellar activity.  For example, \citet{Boyajian2012} found that eclipsing binary M dwarfs obey markedly different mass-\teff and radius-\teff relationships than single field M dwarfs.  \citet{Stassun2012} investigated the effect of chromospheric activity on radius-\teff relationships, due to \teff suppression by the presence of spots.  \citet{Rojas2012} found that two nearby M dwarfs with white dwarf companions are significantly metal-rich compared to the solar neighborhood, leading them to speculate that the M dwarfs were polluted.  These studies imply that the Dartmouth evolutionary isochrones must be used with caution when determining the mass and radius of a binary or active M dwarf based solely on its \teff and metallicity.

In this paper we report accurate stellar parameters for the M dwarf in KOI-256, as well as the co-orbiting white dwarf, measured empirically by combining the ultra-precise {\it Kepler} light curve with ground-based spectroscopic and adaptive optics observations.  In Section \ref{observations}, we describe the observations and the measured parameters used in this analysis, including a revised orbital period and occultation and transit ephemerides.  In Section \ref{Parameters} we combine the measurements to determine the physical parameters of the system.  In Section \ref{Discussion} we discuss the importance of KOI-256 for future studies of PCEBs, white dwarfs and exoplanets.

\section{Observations}\label{observations}

\subsection{Visible-light Adaptive Optics Imaging}

We observed KOI-256 with the Robo-AO laser adaptive
optics and imaging system \citep{Baranec2012} on the Palomar
Observatory 60-inch Telescope on
UT 2012 15 July to look for contaminating sources in the \ik apertures used to contract the light curves. We used a long-pass filter with a 600nm cut-on to
more closely approximate the Kepler bandpass while maintaining
diffraction-limited resolution.
The observation consisted of a sequence of full-frame-transfer
detector readouts at the maximum rate of $\sim$ 8.6 Hz for a total of 120
s of integration time. The individual images
were then combined using post-facto shift-and-add processing using
KOI-256 as the tip-tilt star with 100\% frame selection.
The wavelength response of the Robo-AO visible camera
is similar to the Kepler response, and the image accurately
represents the degree of contamination by other
stars in the Kepler apertures. We achieved high adaptive optics correction
at visible-wavelengths: KOI-256 has a full-width-at-half-
maximum of 0\farcs12. 

\begin{figure}[bh]
\hfill
\begin{minipage}[t]{.45\textwidth}
\begin{center}
\includegraphics[width=3.2in]{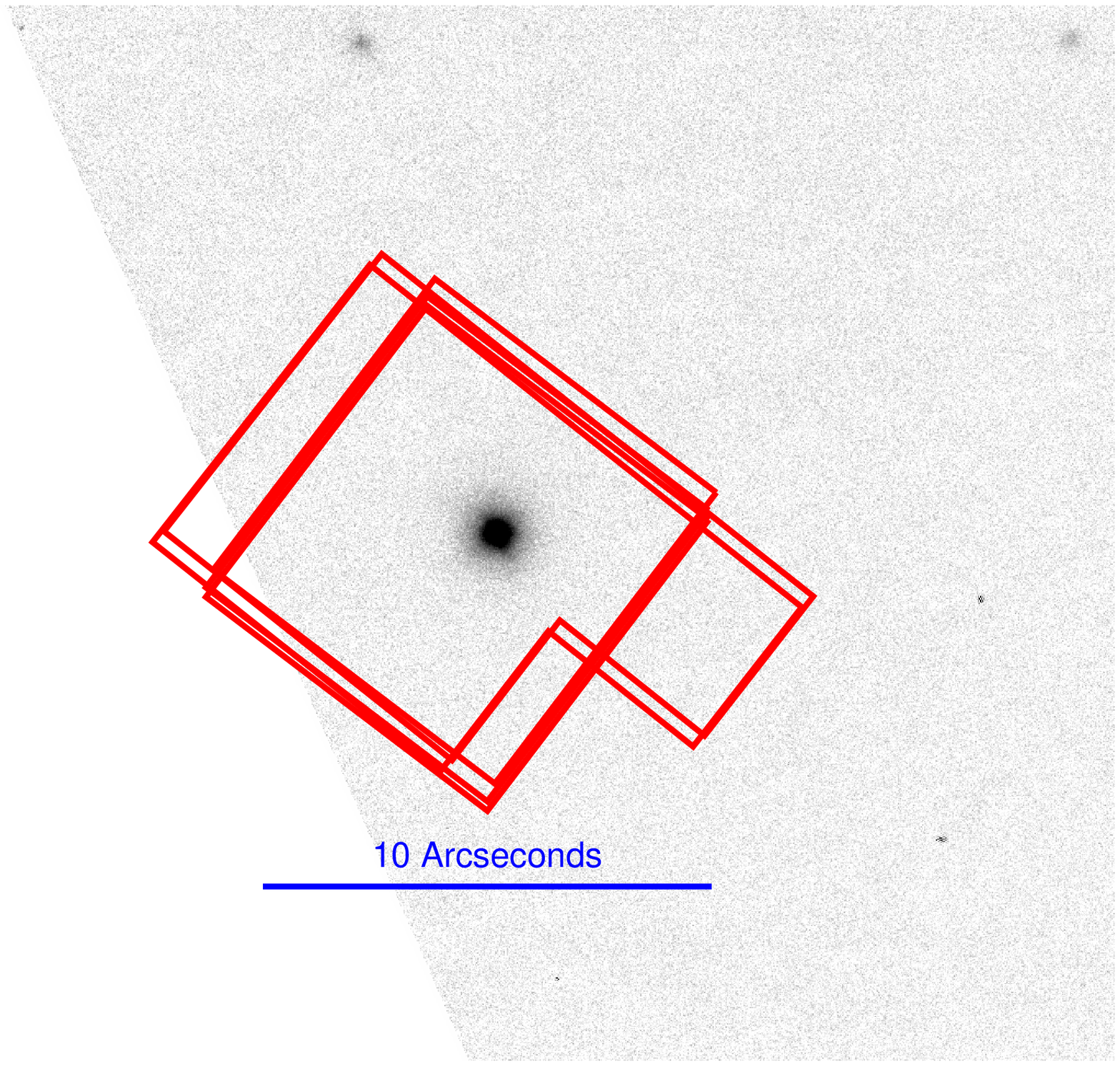}
\caption{Visible-light adaptive optics image of KOI-256 by
the Robo-AO system on the Palomar Observatory 60-inch Tele-
scope. North is up and East is left, and the blue line rep-
resents 10 arcsec. We overlay the \ik apertures ({\it red
boxes}) used to collect short-cadence data. KOI-256 happened
to be near the edge of the Robo-AO field-of-view during integrations,
resulting in the sharp cut-off on the left of the image.
We detect no contaminating stars within the Kepler apertures.\label{image}}
\end{center}
\end{minipage}
\hfill
\begin{minipage}[t]{.45\textwidth}
\begin{center}
\includegraphics[width=3.2in]{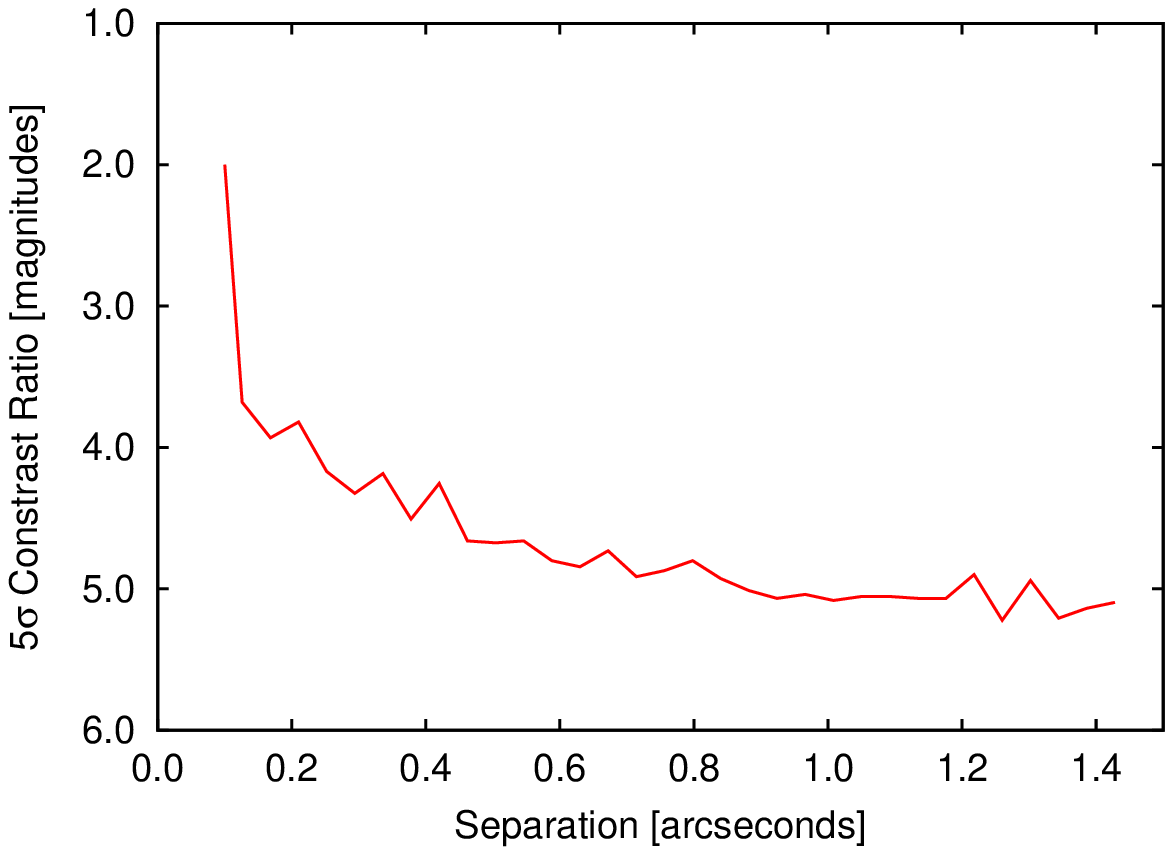}
\caption{Contrast curve showing 5$\sigma$ lower limits on the magnitude difference of a potential background or companion object for a given separation, calculated from the Robo-AO image above.  Based on these data, we proceed assuming all of the flux in the {\it Kepler} light curves come from the M dwarf-white dwarf system.\label{contrast}}
\end{center}
\end{minipage}
\hfill
\end{figure}

Figure \ref{image} shows the final image of KOI-256 and
Figure \ref{contrast} shows a contrast curve indicating the
5$\sigma$ lower-limits on the relative magnitudes of nearby background or contaminating
stars.  The contrast curve was generated through a point spread function (PSF) subtraction algorithm customized for the Palomar Robo-AO imaging system, and based on the Locally Optimized Combination of Images algorithm \citep{Lafreniere2007}. Rather than using PSF reference images obtained sequentially in time on the science target, the algorithm uses reference images derived from the numerous other PSFs in the field of view.  We see no evidence of contamination in the Kepler apertures for objects that are less than 5.0 magnitudes fainter than KOI256 at 1 arcsec of separation. Therefore,
we proceed assuming all of the flux in the Kepler light curve is from the M dwarf-white dwarf system.

\subsection{{\rm Kepler} Light Curve\label{Kepler}}

The {\it Kepler} Mission photometrically monitors targets in two modes: long-cadence mode and short-cadence mode \citep[][]{Koch2010}.  Long-cadence mode involves 30-minute integrations and short-cadence mode involves 1 minute integrations, both taken nearly continuously in the custom {\it Kepler} bandpass spanning roughly 5000 to 9000 \AA.  KOI-256 has publicly available long-cadence mode data from quarters 0 through 11 (2009 May through 2012 January), and short-cadence mode data for quarters 6 and 7 (2010 June through 2010 December).  Both datasets have several gaps in coverage due to quarterly rotation of the spacecraft, data downlink gaps, periods with the spacecraft in safe mode, and intermittent corrections to the spacecraft pointing \citep{Batalha2013}.

\subsubsection{Evidence for Starspots and Synchronous Rotation}\label{avi}

We downloaded the \ik long-cadence data from the Mikulski Archive for Space Telescopes (MAST) for quarters 1, 2, 3, 4, 5, 6, 7, 9, 10, and 11. No data is available for quarters 8 and 12 since during those quarters \koi was located on CCD module 3, which failed permanently about 20 days into quarter 4 \citep{Batalha2013}. To clean the data from systematic trends we used the PyKE software \citep{Still2012} and the co-trending vectors \citep[][Section 2.3]{Barclay2012}, followed by long-term polynomial detrending. 

\begin{figure}[]
\begin{center}
\includegraphics[width=3.0in]{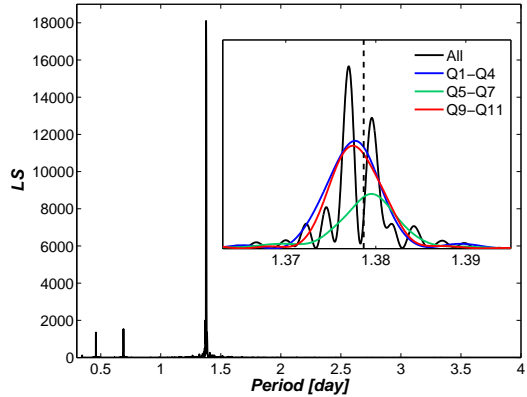}
\caption{Lomb-Scargle median-normalized periodogram of KOI-256 long-cadence data, while ignoring in-occultation and in-transit data. The strongest peak is at 1.38 days, the same period as the orbit, and there are smaller peaks at its harmonics. This shows the light curve has a clear, high signal-to-noise variability with the same period as the orbit. The inset shows a zoom-in on the strongest peak (black solid line), showing it is split into a few lines. The orbital period is marked by a dashed black line. The colored lines in the inset show the periodogram for subsets of Kepler data (see legend). 
\label{per}}
\end{center}
\end{figure}

Figure \ref{per} shows a Lomb-Scargle periodogram \citep{Lomb1976, Scargle1982} of the long-cadence data while ignoring in-occultation and in-transit data. The periodogram's strongest peak is at the same period as the orbit. Figure \ref{per} inset shows a zoomed-in view of the strongest peak, which is in fact split into several peaks. Using the Kepler long-cadence time stamps we estimated the width of a periodogram peak of a fixed period signal spanning the entire Kepler data time span, and found it is similar to the width of the individual sub-peaks. This suggests a slight variation in the photometric period during the course of Kepler monitoring. This is supported by analysis of different subsets of the Kepler data, marked by solid colored lines, where there are small differences in the position of the strongest peaks. The vertical dashed black line marks the 1.38 day orbital period, at the center of the group of peaks. 

Periodic, sinusoidal-like variability at the orbital period can arise from (1) a combination of effects induced by the orbital motion, including beaming, ellipsoidal distortion, reflected light \citep[e.g.,][]{Zucker2007, Shporer2010, Faigler2011, Shporer2011}, or, (2) stellar chromospheric activity of a spin-orbit synchronized star. In the second scenario an active star is rotating at the same period as the orbit, as a result of tidal locking, and the non-uniform longitudinal distribution of spots across the stellar surface results in observed sinusoidal-like variability at the orbital period.

\begin{figure*}[]
\begin{center}
\includegraphics[width=7.0in]{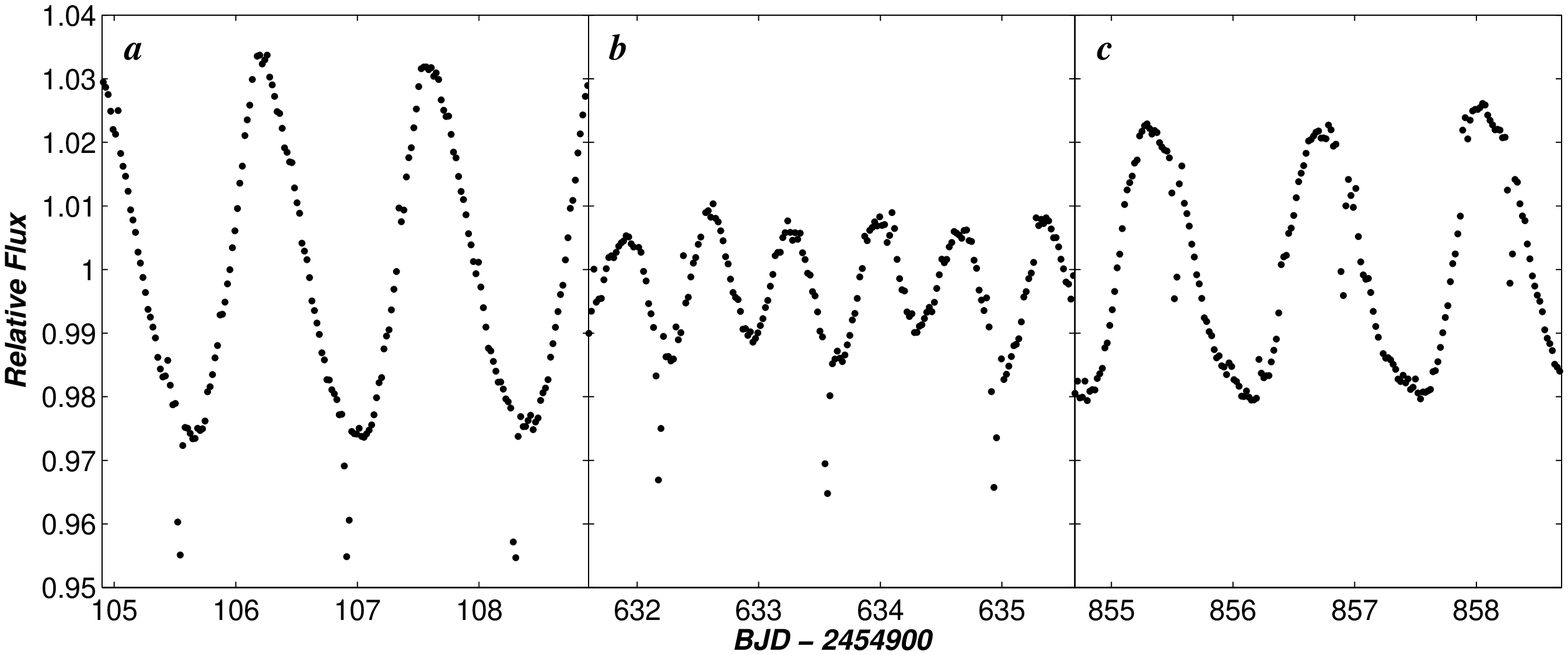}
\caption{Long-cadence {\it Kepler} photometry of KOI-256. Each panel shows four days of relative flux, and panels are separated by several hundred days. All panels show a clear sinusoidal modulation, and the occultation events are seen as a brief drop in relative flux, each consisting of 2--3 long-cadence measurements. The transit events are not apparent in the light curve, and require phase-folding to detect (Section \ref{transit}). The variability pattern shows the typical evolution in amplitude, phase, and overall shape, which is the signature of stellar activity. Compared to panel {\it a}, panel {\it b} shows variability with half the period and a smaller amplitude, and panel {\it c} shows variability at a different phase (note how the occultation is no longer close to time of minimum relative flux), and also a slightly smaller amplitude.  We attribute this behavior to the combination of synchronous rotation of the M dwarf with the orbital period around the M dwarf-white dwarf center of mass, and an evolving spotted surface on the M dwarf which changes over 100 day timescales.\label{long_cadence}}
\end{center}
\end{figure*}

In order to investigate the origin of the photometric variability we visually examined the light curve. Figure \ref{long_cadence} shows three long-cadence light curves each spanning four days, and separated from each other by a few 100 days. The occultations are seen as brief drops in relative flux, spanning 2--3 long-cadence measurements. A visual comparison of the three panels shows an evolution of the light curve shape. The three light curves show different amplitude, and a different relative phase between the occultation and the out-of-eclipse modulation (especially between panels {\it a} and {\it c}). In addition, panel {\it b} shows a double peak modulation, at half the period seen in the other two panels. This evolution in the light curve shape is clear evidence that it originates from spots, since the orbital effects mentioned above should not show evolution at the observed timescales. Moreover, given our estimates of the system parameters (see Section \ref{Parameters}), these orbital effects are expected to show variability of $\sim 0.1\%$, an order of magnitude smaller than the observed amplitude. Finally, the splitting of the periodogram peak in Figure \ref{per} has been observed for other active stars, and could be a sign of differential rotation \citep{Chaplin2013}.

Since the white dwarf brightness is only $\sim2\%$ of the system total brightness in the \ik band, it can not be the origin of the observed photometric variability. Therefore, we conclude that the dM component is spotted and tidally locked to the white dwarf.  In Section \ref{Parameters} we use the fact that the M dwarf is spin-orbit synchronized in combination with the measured \vsini of the M dwarf and the occultation period and duration to empirically constrain the radius of the M dwarf.

We note that the high signal-to-noise ratio (S/N) of the spot-induced variability, and the relatively short rotation period compared to the long time-span of the data, allows for detailed analysis of the spots pattern and their evolution. This can possibly lead to investigation of differential rotation, magnetic activity cycles, and magnetic interaction between the two stars.  

We also note that beyond the eclipses and quasi-sinusoidal variability, the light curve shows also many flares, with amplitudes that vary from $<$1\% to $\sim$10\%, which occur with a frequency that is comparable to the orbital frequency.

\subsubsection{Precise Orbital Period}\label{period}

\begin{figure}[]
\begin{center}
\includegraphics[height=3.0in,angle=90]{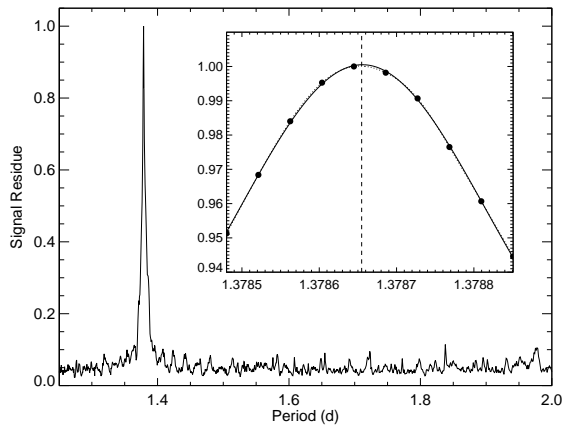}
\caption{Signal residue vs. occultation period from the BLS analysis for low-resolution and high-resolution near the peak({\it insert}).  We use two methods to find the peak of the high-resolution power spectrum: a Gaussian-fit and spline-interpolation.  Averaging the methods gives an occultation period of 1.3786548 $\pm$ 0.00001 days.\label{bls}}
\end{center}
\end{figure}

We downloaded short-cadence {\it Kepler} light curve data of KOI-256 from MAST for quarters 6 and 7, spanning roughly 6 months nearly continuously.  First, we measured the orbital period of KOI-256 using the Box-fitting Least Squares algorithm \citep[BLS,][]{Kovacs2002} applied to the short cadence data in an iterative scheme. For the first iteration, the parameters of the BLS algorithm were allowed to vary considerably around the nominal transit parameters from \citet{Batalha2013}. Successive iterations of BLS were run over smaller regions of frequency space, sampling the signal residue in finer detail and placing tighter constraints on the transit duration. Our final iteration sampled the frequency range of 0.7247 to 0.7260 $day^{-1}$ sixty times using 500 bins across the phase folded data with the transit duration constrained to be within 0.031 and 0.032 times the period. The peak of the signal residue is estimated with two methods: fitting a Gaussian model and using a cubic spline interpolation between BLS samples. These methods return values of 1.3786570 and 1.3786526 days, respectively, in agreement at the level of $4\times 10^{-6}$, and shorter than the \citep{Batalha2013} value by about 2 s. We take the mean of the two measurements as the final value, and we estimate the uncertainty to be $1\times 10^{-5}$ days, or roughly 0.8 s.  Figure \ref{bls} shows the signal residue from the BLS analysis for a coarse frequency resolution, with the final high resolution analysis inlaid. The Gaussian fit and the cubic spline interpolation are shown as solid and dotted lines, respectively, with the final value of 1.3786548 days designated with the dashed line. 

\subsubsection{Occultation Light Curve}\label{occultation}

With an accurate period for the occultation events, we phase-folded the short-cadence data to fit an occultation model to the light curve.  We used short-cadence data for the occultation fit, and neglected the long-cadence data, since high-cadence is required to resolve the ingress and egress.  First, we detrended the light curve to remove the strong variations due to the aforementioned rotational modulation of the M dwarf and occasional flares.  For each occultation event, we focused on a 3.36 hr wide window roughly centered around the occultation.  We applied a running median-filter to the out-of-occultation data points using a box size of 14 minutes.  We fitted a second-order polynomial to the median-filtered, out-of-occultation data, then divided all of the original data points in the 3.36 hr window by the polynomial fit, including the data taken during occultation.

\begin{figure}[]
\begin{center}
\includegraphics[width=3.3in]{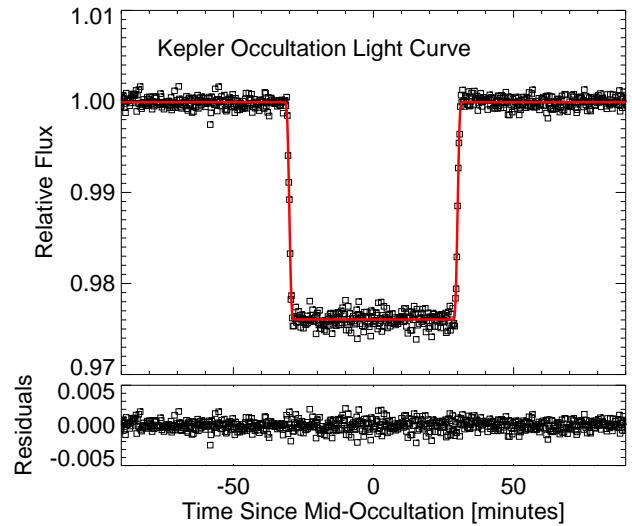}
\caption{Phase-folded {\it Kepler} short-cadence light curve of KOI-256 during the occultation of the white dwarf as it passes behind the M dwarf. We have removed the sinusoidal modulation by fitting a polynomial to the out-of-eclipse data, normalized the resulting values, phased the data to match the period as measured using the BLS algorithm.  and binned the data to 15 s windows.  We fitted a model described in Section \ref{occultation} to the un-binned light curve; however, we binned the data into 15 s windows in the figure for clarity.  We show the residuals in the lower panel.\label{occultation_plot}}
\end{center}
\end{figure}

We then phase-folded the de-trended data using the revised period and binned the data into 15 s windows.  For each bin we removed 3$\sigma$ outliers and calculated the mean of the data points within the bin.  Figure \ref{occultation_plot} shows the resulting de-trended and phase-folded short-cadence occultation light curve.

We fitted a custom occultation model, which unlike planet transit models \citep[e.g.][]{Mandel2002} leaves the ingress/egress times independent of the occultation depth. We assumed that the radius of the white dwarf was much smaller than that of the M dwarf, and modeled the shape of the occultation as a circle with a linear limb-darkening coefficient passing behind a flat edge, with the occultation duration and ingress/egress durations independent parameters. We convolved the modeled light curve with a boxcar window of length 58.8 s to simulate the effects of the 58.8 s Kepler short cadence integrations. We fitted the model to the binned and phase folded data using a Levenberg-Marquardt algorithm.  Varying the linear-limb darkening coefficient for the white dwarf had little effect on the resulting depth, duration or ingress/egress time relative to their estimated uncertainties; therefore, we chose to assume no limb-darkening in the white dwarf.

The fitted parameters and corresponding 1$\sigma$ uncertainties are summarized in Table \ref{parameters_table}.  The fitted fractional occultation depth is 0.0238743 $\pm$ 0.000047, similar to the value listed for the planet-candidate signal in \citet{Batalha2013}, but with higher precision.  However, the fitted occultation duration is is 60.093 $\pm$ 0.037 minutes, and ingress and egress duration is 1.535 $\pm$    0.067 minutes.  If the occultation were in fact a transiting planet, it would have to be Jupiter-sized in order to cause such a large fractional change.  A transiting Jupiter-sized planet would result in a much larger ingress-to-occultation duration ratio: of order 1/10.  The measured ingress-to-occultation duration ratio of $\sim$1/40 is instead consistent with an Earth-sized emitting object passing behind the M dwarf.  In the following sections we support this interpretation with a detection of the transit of the white dwarf in front of the M dwarf, an undetected infrared occultation observation and radial velocity observations of the M dwarf.

\begin{figure}[]
\begin{center}
\includegraphics[width=3.3in]{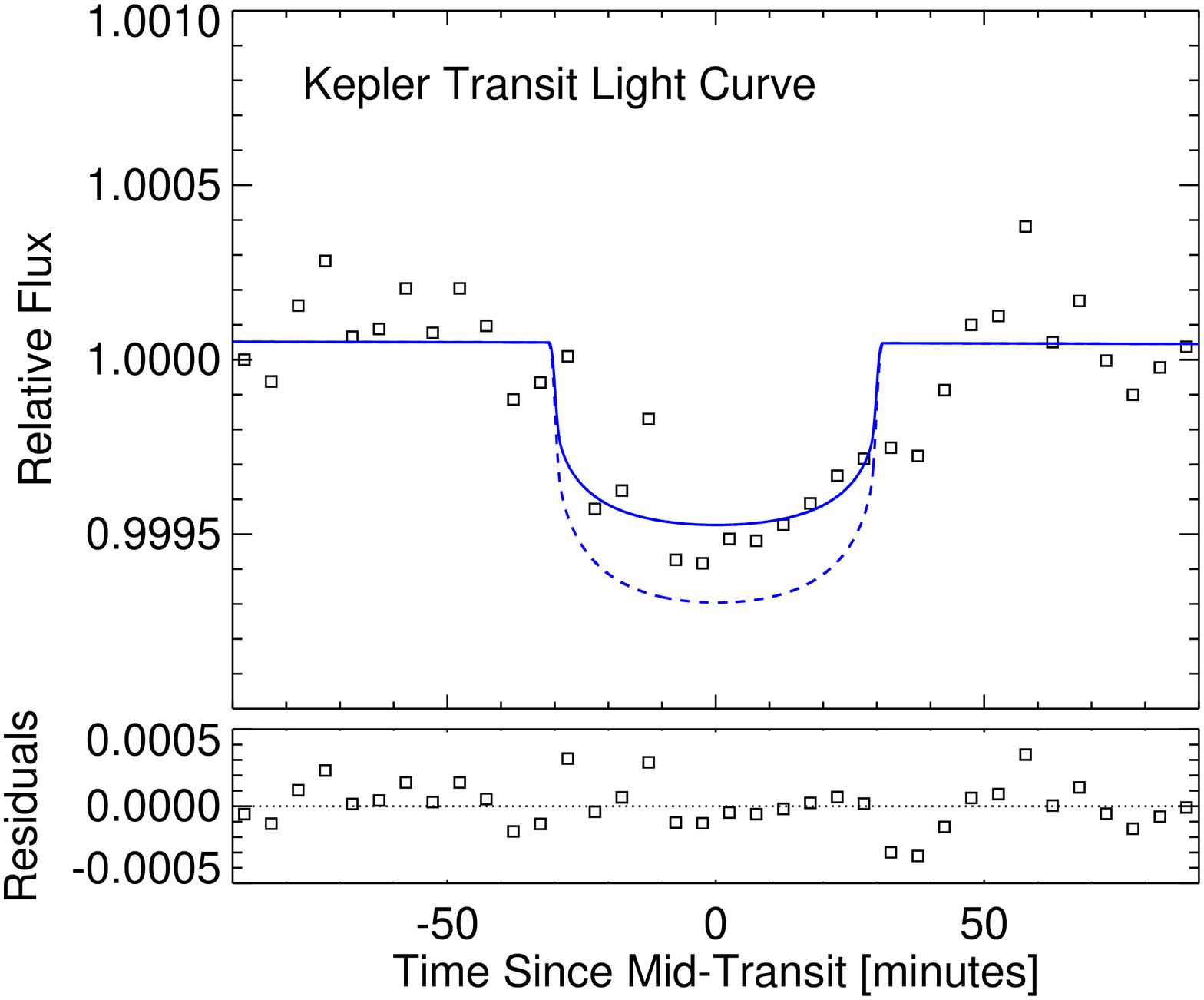}
\caption{Phase-folded {\it Kepler} short-cadence light curve of KOI-256 during the transit of the white dwarf in front of the M dwarf.  {\it Solid line:} best fitting model light curve.  {\it Dashed line:} the model light curve expected without gravitational lensing.  We applied the same detrending method used for the occultation events.  We fitted a conventional planetary-transit model with limb-darkening from \citet{Mandel2002} to the un-binned short-cadence light curve; however, we binned the data into 5-minute-wide windows in the figure for clarity.  Since the M dwarf is significantly larger than the expected Einstein radius around the white dwarf during transit, the light curve follows the same shape as a planetary transit \citep{Agol2003}.\label{transit_plot}}
\end{center}
\end{figure}

\subsubsection{Transit Light Curve}\label{transit}

We used a similar procedure to analyze the {\it transit} of the white dwarf in front of the M dwarf.  Unlike the occultation events, the transit events are not immediately apparent in the raw light curve due to the small geometric size of the white dwarf compared to the M dwarf.  We de-trended and phase-folded the short-cadence light curve using the same procedure used in the previous section, except centered around the expected transit times assuming zero eccentricity: halfway between the occultation events.

\citet{Agol2003} showed that, for transits of compact objects where the Einstein radius of the lens is significantly less than the radius of the source, the light curve obeys the same shape as a non-lensed planetary transit, but with a shallower or inverted transit depth.  This is true even when limb-darkening of the source is included.  This allows us to fit a conventional un-lensed transit light curve to the white dwarf transit data using an {\it effective} radius ratio parameter $(R_1 / R_2)$, which relates to the stars' physical radii and Einstein radius as such:

\begin{equation}\label{lensing}
\Big({R_1 \over R_2}\Big)^2 =  { R_{\rm WD}^2 - 2 R_{\rm Ein}^2 \over R_{\rm dM}^2 }
\end{equation}

\noindent where $R_{\rm WD}$ is the radius of the white dwarf, $R_{\rm dM}$ is the radius of the M dwarf and $R_{\rm Ein}$ is the Einstein radius.  Assuming the source-to-lens distance is much smaller than the lens-to-observer distance, the Einstein radius takes the following form:

\begin{equation}
R_{\rm Ein} =  \sqrt{ 4 \, G \, M_{\rm WD} \, a \over c^2 }
\end{equation}

\noindent where $M_{\rm WD}$ is the mass of the white dwarf, $a$ is the combined semi-major axis of the binary, $G$ is Newton's gravitational constant and $c$ is the speed of light.  

After subtracting the photospheric flux-contribution from the white dwarf using the occultation depth, we fit the  transit model of \citet{Mandel2002} to the data, using the expected quadratic limb-darkening coefficients of the M dwarf calculated by  \citet{Claret2011}.  We fixed the ingress/egress time and transit duration to match the values measured for the occultation, and fixed the mid-transit time to halfway in between the occultation events.  The only free parameters in the fit are the effective radius ratio $(R_1 / R_2)$ and a slope and offset for the out-of-eclipse light curve.  We measure a combined effective radius ratio of 0.0218 $\pm$ 0.0008.  In Section \ref{masses}, we use the effective radius ratio in combination with other measurements to determine the physical stellar masses and radii and the system orbital parameters.

\subsection{Infrared Occultation Light-Curve}\label{wirc}

We observed an occultation event of KOI-256 with the Wide-Field Infrared Camera (WIRC) on the Palomar 200-inch Hale Telescope \citep{Wilson2003} on UT 2011 August 19 from 04:50:53 UT to 08:12:29 UT.  If the occultation were in fact a transiting planet, we would expect the same fractional depth in the infrared as in the \ik bandpass.  We used the $H$-band filter centered on 1.635 $\mu$m and spanning $\pm$ 0.2 $\mu$m.   We took 618 10 s exposures over the course of 202 minutes, corresponding to a duty cycle of 51\%. We heavily defocused the telescope to spread out starlight and mitigate flat fielding errors, and ``stared" on the target for the whole period.  The airmass changed from 1.055 to 1.029 over the course of the observation. 

Reduction of the WIRC data was carried out using the routines developed in \citet{Zhao2012}.  We first subtracted the science images with corresponding averaged dark frames and a scaled ``super-sky" image. The ``super-sky" image was constructed with sky background frames taken right before and after the science frames. The data were then normalized with a master flat field. Thirteen reference stars within the flux range of 0.38 - 1.1 times that of KOI-256 were selected to correct for the correlated common-mode systematics in the light curve. Aperture photometry was then performed on the target and the reference stars, following the steps of \citet{Zhao2012}.  A final aperture with radius of 16.5 pixels (4\farcs125) gave the smallest scatter in the data and was used for the photometry. A sky annulus with a 25 pixel inner radius and 35 pixel outer radius was used for background estimation. Both the reduced out-of-occultation and in-occultation data closely follow the Gaussian noise expectations when binned (the scatter reduces as 1/$\sqrt{N}$, where $N$ is the number of binning points), suggesting most of the correlated noise was corrected in the light curve. The final precision of the fitted depth of the light curve is 660 ppm, limited by the low brightness of the star and the systematics stemming from large centroid drifts ($>$ 15 pixels) over the course of the observation. 

\begin{figure}[]
\begin{center}
\includegraphics[width=3.3in]{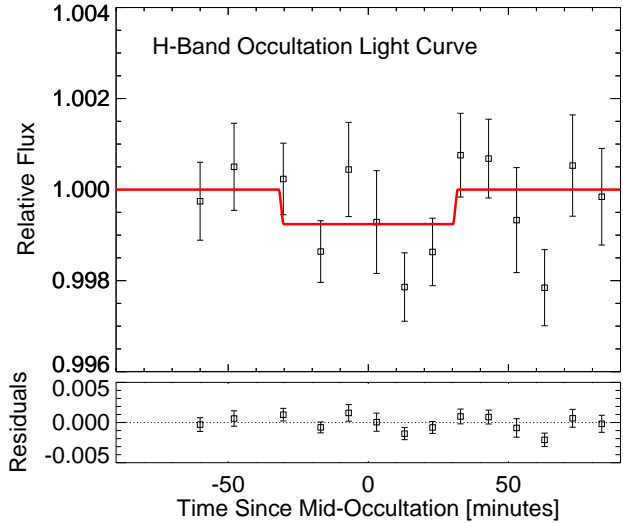}
\caption{WIRC $H$-Band light curve for a single occultation of the white dwarf by the M dwarf in KOI-256.  We set a 3$\sigma$ upper limit of 0.00275 for the fractional occultation depth in $H$-band.  The lack of a deep transit as seen in the {\it Kepler} data is consistent with the white dwarf interpretation of the system.  In $H$-band, the expected flux ratio for the white dwarf is roughly 0.001, using the M dwarf and white dwarf temperatures and radius ratio determined in Section \ref{teff}.\label{wirc_plot}}
\end{center}
\end{figure}

To determine the occultation depth, we fixed the timing and duration to values determined in from the {\it Kepler} light curve and only allowed the depth to change in the fit. The UTC mid-exposure time of each images was converted to BJD$_{TDB}$ using the UTC2BJD routine by \citet{Eastman2010}.  The best-fit gives an occultation depth of 0.00076 $\pm$ 0.00066, with a 3$\sigma$ upper limit depth of 0.00275.  The uncertainty was determined using the largest value from the Levenberg-Marquardt fit, bootstrap, and residual permutation.  Figure \ref{wirc_plot} shows the best-fit light curve model and the normalized data. 

The occultation depth in $H$-band is significantly smaller than the occultation depth in the \ik band, consistent with the white dwarf occultation interpretation.  In Section \ref{Parameters}, we calculate the stellar radii and white dwarf effective temperature (values listed in Table \ref{parameters_table}).  For the radii and temperatures we measure, we expect an occultation depth of 0.001 in $H$-band, consistent with our upper limit, owing to the large flux contrast between the stars at this wavelength.  

\subsection{Near-infrared Radial Velocities}\label{rvs}

We observed KOI-256 with the TripleSpec Spectrograph on the Palomar 200-inch Hale Telescope on UT 2012 August 5, 6, 7, 8 and 11 for the purpose of measuring radial velocities of the M dwarf in KOI-256.  TripleSpec is a near-infrared, long-slit spectrograph covering 1.0 to 2.5 $\mu$m simultaneously with a resolving power of 2700 \citep{Herter2008, Wilson2004}.  We acquired spectra of KOI-256 nine times over the course of five orbital periods to measure the object's radial velocity variation.   Each measurement consisted of four 80 s exposures, each taken at two positions along the TripleSpec slit in order to subtract and remove contaminating sky emission lines.  We reduced the spectra using the {\tt SpexTool} program modified for TripleSpec at Palomar \citep[][M. Cushing 2011, {\it private communication}]{Cushing2004}.  For each observation of KOI-256, we acquired a spectrum of a nearby A0V star (HD 179095) to calibrate and remove telluric absorption lines using the {\tt xtellcor} package \citep{Vacca2003}.  Each resulting spectrum of KOI-256 had median signal-to-noise ratio of roughly 100 per pixel in $K$-band.

The modified {\tt SpexTool} program calculates a wavelength solution for reduced spectra using sky emission lines.  We acquired deep (300 sec) exposures of the sky each night we observed KOI-256 to provide an accurate wavelength solution for each spectrum.  We also observed Gl\,87 on UT 2012 August 6 using the same observing strategy, and achieved a similar S/N.  Gl\,87 is a nearby \citep[dis=10.4 pc,][]{vanleeuwen2007} M2.5 dwarf \citep{Reid1995} with an absolute radial velocity of 7.0 $\pm$ 2.0 $km \, s^{-1}$ \citep{Wilson1953, Joy1947}.  We measured an effective temperature of 3518 $\pm$ 50 K for Gl\,87 using the K-band indices of \citet{Rojas2012}.  The effective temperature is similar to what was measured for KOI-256 using the same indices \citep[\teff = $3450 \pm 50\,K$,][]{Muirhead2012b}.  The similarity between the $K$-band spectra of the stars makes Gl\,87 a suitable target for use as a calibrated absolute radial velocity template spectrum for KOI-256.

To measure radial velocities of the M dwarf component of KOI-256, we use a select region of $K$-band from 2.2 to 2.4 $\mu$m, where M dwarfs have strong Na, Ca and CO features and the flux contribution from the white dwarf is negligible.  First, we cross-correlated the spectrum of Gl 87 with each spectrum of KOI-256.  The resulting radial velocity from the cross-correlation was used as a starting value for a Levenberg-Marquardt fit with three free parameters: two to adjust the continuum of the Gl 87 spectrum to match KOI-256 and one to apply a radial velocity shift to the template. The Levenberg-Marquardt fit result was further refined and the posterior probability distribution explored using a Markov Chain Monte Carlo algorithm. The posterior probability distributions provide more robust measurements of the true radial velocity shifts, as the Levenberg-Marquardt fit is susceptible to local maxima in the distribution. The posterior probability distribution also provides uncertainty estimates for the measured radial velocities. The typical 1$\sigma$ uncertainty on the measurements is 4 km ${\rm s^{-1}}$.

\begin{figure}[]
\begin{center}
\includegraphics[width=3.3in]{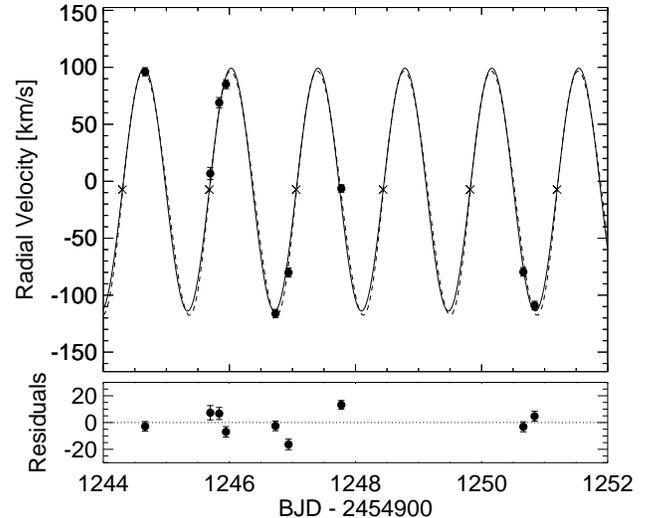}
\caption{Radial velocities of the M dwarf in KOI-256 measured with TripleSpec and using a spectrum of Gl 87 as a template and absolute RV reference, corrected for the motion of the observatory relative to the solar system barycenter.  We fitted two Keplerian models to the data using the {\tt rvlin} package \citep{Wright2009}: one with the eccentricity fixed to zero ({\it solid line}) and one with the eccentricity as a free parameter ({\it dashed line}, e = $0.056$).  We do not see a significant improvement in the fit; therefore we proceed assuming zero eccentricity.  We fixed the orbital period and phase based on the occultation period and ephemeris as measured using {\it Kepler} light curve (see Section \ref{period}, the occultation ephemeris times are marked with x's).  The residuals are plotted for the zero-eccentricity model.  The root-mean-square of the residuals is 8.4 $\rm km \, s^{-1}$, which is roughly twice the estimated per measurement uncertainty.\label{rv_plot}}
\end{center}
\end{figure}

After correcting for the motion of the observatory relative to the solar system barycenter and the archival radial velocity measurement of Gl 87, we fitted a Keplerian model to the radial velocity measurements of KO-256 using the {\tt rvlin} package \citep{Wright2009}.  We fixed the orbital period and phase to that expected from the occultation events in the {\it Kepler} light curve.  Figure \ref{rv_plot} shows our measured radial velocities and the Keplerian fit and Table \ref{rv_table} shows the best-fitting Keplerian parameters.  The large amplitude and phase of the RV variations are consistent with a stellar-mass companion passing behind the M dwarf (occultation), rather than a planetary mass companion passing in front of the M dwarf (transit, $180^{\circ}$ phase difference).  When combined with the trapezoidal shape and sharp ingress and egress of the occultation light curve (Section \ref{occultation}), this provides strong evidence for the M dwarf-white dwarf interpretation of KOI-256. 

\begin{table}[]
\begin{center}
\caption{TripleSpec Radial Velocities of the M Dwarf in KOI-256 and Keplerian Fit\label{rv_table}} 
\begin{tabular}{l c} 
\hline\hline                        
BJD  - 2454900 & {RV (km $\rm s^{-1}$)} \\ [0.5ex] 
\hline                  

       1244.6636 &    96.1 $\pm$   3.6 \\
      1245.6961 &     6.8 $\pm$   5.4 \\
      1245.8384 &    69.0 $\pm$   4.5 \\
      1245.9457 &    85.1 $\pm$   3.9 \\
      1246.7308 &  -116.1 $\pm$   3.6 \\
      1246.9384 &   -80.1 $\pm$   3.9 \\
      1247.7754 &    -6.4 $\pm$   3.3 \\
      1250.6623 &   -79.6 $\pm$   3.7 \\
      1250.8408 &  -109.0 $\pm$   3.7 \\

\hline\hline                        
\\
Parameter & Value \\ [0.5ex] 
\hline                  
$Per$ (days, fixed) & 1.3786548 \tablenotemark{a}\\
$T_p$ (BJD, fixed) & 2455374.324838\tablenotemark{a}\\
$e$ (fixed) & 0.0\\
$\omega$ (fixed) & 90.$^{\circ}$0\\
$K$ (km $\rm s^{-1}$) &  105.9 $\pm$ 1.6 \\
$\gamma$ (km $\rm s^{-1}$) & -7.3 $\pm$ 2.0\tablenotemark{b}\\
$dv / dt$ (fixed) & 0.0 \\
\hline 
\tablenotetext{a}{Fixed to the occultation period and ephemeris measured from the {\it Kepler} light curve (see Section \ref{Kepler}).}
\tablenotetext{b}{Measured relative to Gl 87, assuming literature value of 7.0 $\pm$ 2.0 km $\rm s^{-1}$ \citep{Wilson1953, Joy1947}.}
\end{tabular}
\end{center}
\end{table}

\subsection{High-resolution Optical Spectroscopy}\label{hires}

We observed the optical to far-red (3640-7820 \AA) spectrum of KOI-256 using the HIgh-Resolution Echelle Spectrometer \citep[HIRES][]{Vogt1994}. We used the standard observing setup of the California Planet Survey \citep[CPS;][]{Howard2010, Johnson2010} including the red cross-disperser and the C2, 0\farcs86 wide decker, with the iodine cell out of the light path.  Our 500 s exposure resulted in an S/N of roughly 13 at 5500 \AA. 

\subsubsection{\vsini of the M dwarf}\label{vsini_section}

\begin{figure*}[]
\begin{center}
\includegraphics[width=6.5in]{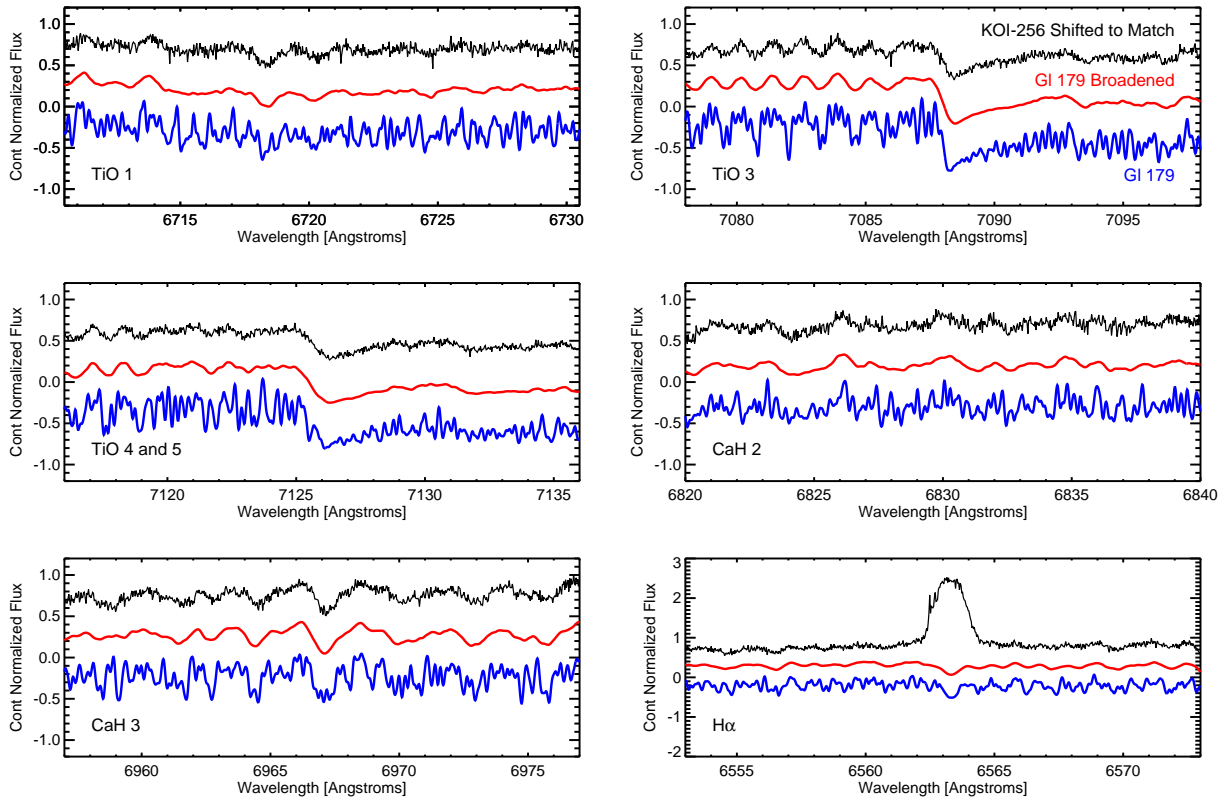}
\caption{HIRES spectra of KOI-256 ({\it black}), Gl 179 ({\it blue}) and a rotationally broadened version of Gl 179 (broadened to $V\sin i$ = 19.6 km $\rm s^{-1}$, {\it red}) for select PMSU molecular bands \citep{Reid1995} including H$\alpha$.  The spectra are continuum normalized and manually shifted in wavelength to overlap within each spectral region.  Gl 179 has similar \teff and [M/H] to the M dwarf component of KOI-256, as measured by the \citet{Rojas2012} methods.  The KOI-256 spectra are dominated by the M dwarf component, which constitutes nearly 98\% of the flux in these regions based on the occultation depth.  The M dwarf in KOI-256 is significantly rotationally broadened compared to to Gl 179, a known slow rotator \citep{Browning2010}.  We measure the $V\sin i$ of the M dwarf to be 19.6 $\pm$ 1.0 km $\rm s^{-1}$ (see Figure \ref{vsini_plot}).  We note the strong H$\alpha$ emission in the M dwarf, for which we measure an equivalent width of 3.32 $\pm$ 0.05 \AA.  The H$\alpha$ emission likely comes from chromospheric activity in the rapidly rotating M dwarf, consistent with the rotation-activity correlation in M dwarfs \citep{Morgan2012}.\label{hires_plot}}
\end{center}
\end{figure*}

Figure \ref{hires_plot} shows the HIRES spectra of KOI-256 and a nearby M dwarf monitored by CPS, Gl 179, in molecular bands corresponding to select Palomar Michigan State University (PMSU) indices including H$\alpha$ \citep[][]{Reid1995}.  Gl 179 is an M3 Dwarf with a measured effective temperature \teff=3424 $\pm$ 16 K and overall metallicity [M/H] = +0.17 $\pm$ 0.12 dex as measured by \citet{Rojas2012}.  \citet{Muirhead2012b} used the same $K$-band spectroscopic techniques to measure similar values for KOI-256 (\teff = $3450 \pm 50 K$, $\rm M/H = +0.31\pm0.10$).  \citet{Browning2010} measured the rotational broadening of Gl 179 to be \vsini$ < 2.5 \, {\rm km s^{-1}}$ using archival CPS data.  With such a low \vsini and similar \teff, Gl 179 is ideal for use as a template for measuring the \vsini of KOI-256.  We combined all of the spectra taken of Gl 179 over the course of the CPS survey using regions that are not contaminated by molecular iodine absorption.  The result is a spectrum with an S/N of roughly 2500 per pixel.  The large radial velocity change of the M dwarf as it orbits the white dwarf results in a smearing of the HIRES spectrum during the exposure.  However, when added in quadrature to the rotational broadening, the smearing effect contributes less than our estimated \vsini measurement errors.

To measure the \vsini of KOI-256, we following the approach of \citet{Browning2010}.  For each spectral region in Figure \ref{hires_plot} (excluding H$\alpha$) we cross-correlated the high S/N spectrum of Gl 179 with a rotationally broadened version of itself for a grid of \vsini values spanning 1 to 30 \kms.  We then fit a Gaussian profile to the resulting cross-correlation function (XCFs), and recorded the full-width-at-half maximum: the XCF FWHM.  With the resulting calibration between XCF FWHM and \vsini, we cross-correlated the unbroadened Gl 179 spectrum with the KOI-256 spectrum, and recorded the XCF FWHM for each spectral region.  We then interpolated the XCF FWHM onto the calibration XCF FWHM versus \vsini relation.  

\begin{figure}[]
\begin{center}
\includegraphics[width=3.3in]{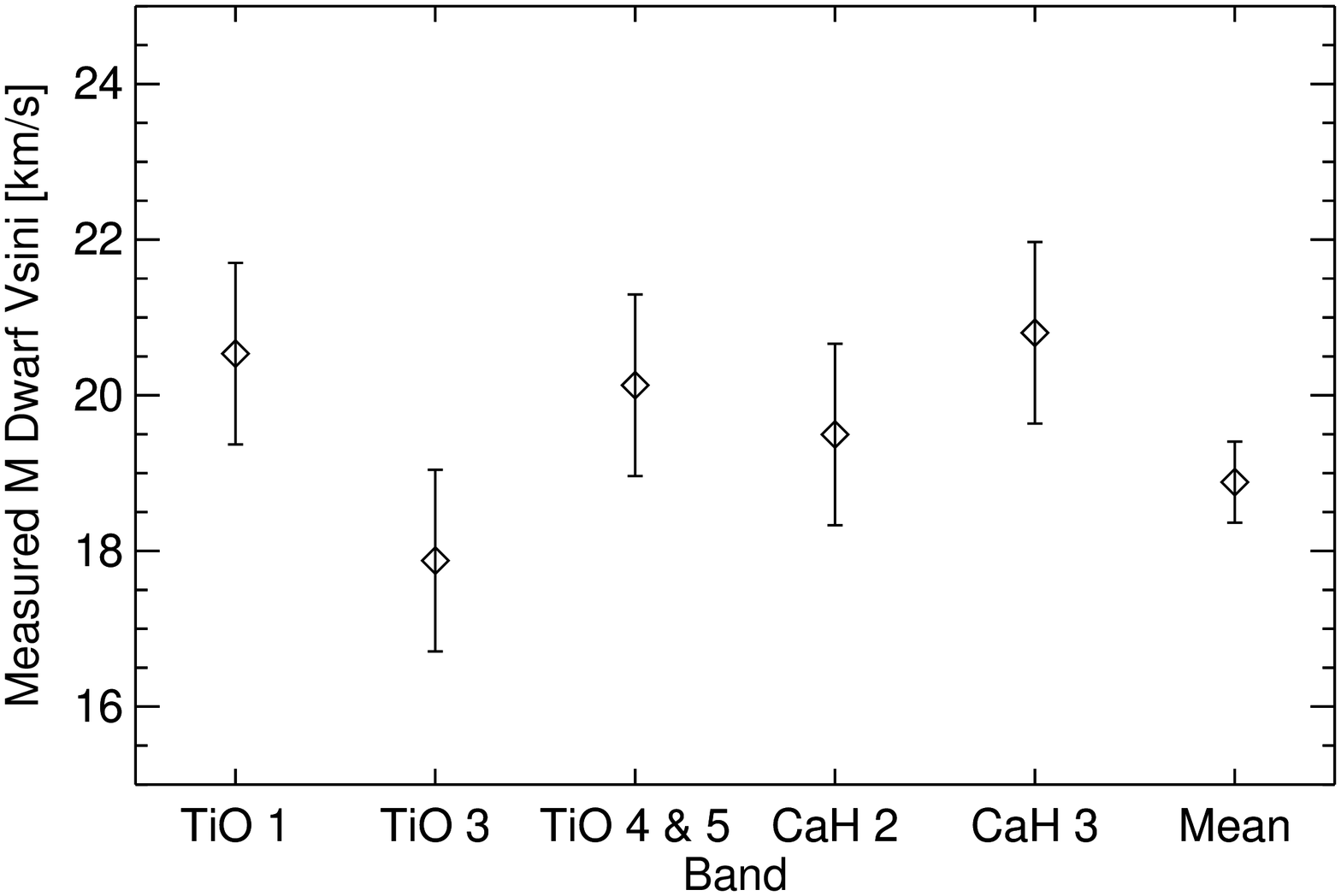}
\caption{Plot showing the measured \vsini of the M dwarf in KOI-256 using different feature-rich regions of the HIRES spectrum corresponding to the bands designated TiO 1, TiO 3, TiO 4, CaH 2 and CaH 3 by \citet[][]{Reid1995}, as well as the mean from all regions.  We follow the procedure of \citet{Browning2010} to measure \vsini, using Gl 179 as a slowly-rotating template spectrum (see Section \ref{hires}).  We take the weighted mean of the measurements as our best estimate (19.67 km $\rm s^{-1}$) with an uncertainty of 0.52 km $\rm s^{-1}$.\label{vsini_plot}}
\end{center}
\end{figure}

The procedure requires an assumption about the linear limb-darkening coefficient for the broadening kernel used on the Gl 179 template spectrum.  We use the value 0.6715 calculated by \citet{Claret2011}, corresponding to the limb-darkening expected from an M dwarf with \teff=3400 $K$ and $log(g)=5.0$.  We take the weighted mean of the measurements for the bands as our best estimate (19.67 km $\rm s^{-1}$) with an uncertainty of 0.52 km $\rm s^{-1}$.  Figure \ref{vsini_plot} illustrates the consistency of the measured \vsini for each of the different bands and the weighted mean.

\subsubsection{Emission Line Strengths}\label{halpha}

We note that KOI-256 shows strong Balmer emission lines and \caii H \& K.  Table \ref{hires_table} lists equivalent widths of selected emission lines as measured in the HIRES spectrum, and Figure \ref{hires_plot} shows the strong H$\alpha$ emission relative to Gl 179.   We also report the line and continuum spectral regions used to calculate the equivalent widths.  We calculated the uncertainty in the equivalent widths by injecting noise into the spectra and recalculating.  The emission likely comes from chromospheric activity in the M dwarf related to rapid rotation.  This is consistent with the rotation-activity correlation for PCEB M dwarfs measured by \citet{Morgan2012} using spectra from the Sloan Digital Sky Survey.

\begin{table}
\begin{center}
\caption{Equivalent Width Measurements for KOI-256\label{hires_table}} 
\begin{tabular}{l c c c} 
\hline\hline                    

Line & EW (\AA) & Line Region (\AA) & Continuum (\AA) \\ [0.5ex] 
\hline                  
H$\alpha$ &  3.40 $\pm$  0.06 & 6561-6565 & 6557-6560, 6566-6569 \\
H$\gamma$ &  3.87 $\pm$  0.26 & 4339-4342 & 4335-4338, 4344-4347 \\
H$\delta$ &  3.67 $\pm$  0.42 & 4100-4104 & 4096-4099, 4105-4108 \\
H$\epsilon$ &  3.97 $\pm$  0.21 &  3969.5-3971 & 3971-3975 \\
H$\zeta$ &  6.02 $\pm$  0.48 &  3887-3890 & 3883-3886, 3892-3895 \\
\caii H & 16.28 $\pm$  3.33 &  3968-3969.5 & 3971-3975\\
\caii K & 13.56 $\pm$  0.46 &  3932-3934 & 3928-3930, 3936-3939 \\
\hline
\end{tabular}
\end{center}
\end{table}

\subsection{Low-resolution Optical Spectroscopy}\label{esi_section}

We observed KOI-256 with the Echellette Spectrograph and Imager (ESI) on the Keck II Telescope \citep{Sheinis2002} on UT 2012 October 9, for the purpose of detecting the contribution to the spectrum from the white dwarf companion at blue wavelengths.  We operated ESI in ``low-D" mode, using a 1\farcs0 slit to obtain a single-order spectrum from 3900 \AA to 11000 \AA.  The resolving power of ESI in low-D mode varies from R=6000 at 3900 \AA to R=1000 at 11000 \AA.  We co-added 5 separate 300 s exposures, so as to acquire enough flux at the blue end of the detector without saturating the red end, which contains significant flux from the M dwarf component.

We also observed the spectro-photometric standard BD+28 421 using identical settings but with a single exposure of 10 s of integration.\footnote{\url{http://www.eso.org/sci/observing/tools/standards/spectra.html}}  After subtracting a bias frame from the raw images and dividing by a high signal-to-noise twilight flat, we extracted both stars' spectra using the ATV image display software \citep{Barth2001}.  We fitted a custom wavelength solution to the extracted spectra based on the positions of selected, bright HgXe emission lines in a calibration exposure.  Figure \ref{esi} shows the flux-calibrated ESI spectrum.  The optical spectrum is consistent with an M3 dwarf star, and we include a template spectrum from \citet[][]{Pickles1998} in the figure.  We do not, however, detect any flux from the white dwarf in the blue part of the spectrum.  In Section \ref{teff}, we convert the M3 template spectrum to a surface brightness using an empirical radius measurement of another M3 star, Gl 581, for the purpose of determining the white dwarf effective temperature.  We show that the white dwarf is relatively cool and therefore should not appear in the ESI spectrum.

\subsection{GALEX NUV Photometry}

\begin{figure*}[]
\begin{center}
\includegraphics[width=6.6in]{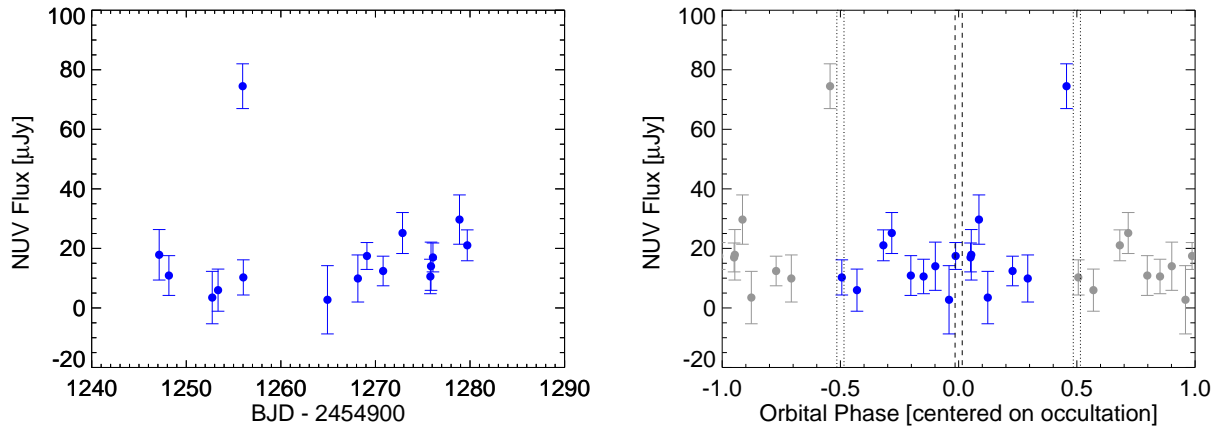}
\caption{GALEX NUV flux from KOI-256.  {\it Left}: flux in $\mu Jy$ vs. barycentric Julian date (BJD) of the observation.  {\it Right}: same, but phase folded to the occultation period with the occultation ingress and egress ephemera represented by {\it dashed lines} and the transit by {\it dotted lines}.  Given the strong H$\alpha$ emission in the M dwarf and low white dwarf temperature (see Section \ref{teff}), we ascribe the NUV emission to the M dwarf chromosphere and not the white dwarf photosphere.\label{galex}}
\end{center}
\end{figure*}

KOI-256 was observed with the near-ultraviolet (NUV) channel on the {\it Galaxy Evolution Explorer} \citep[GALEX,][]{Martin2003} spacecraft on UT 2012 August 7, 8, 13, 16, 25, 28, 29 and 31, and 2012 September 02, 05, 08 and 09 as part of an independent program to mosaic the {\it Kepler} field (J. Lloyd 2012, {\it private communication}).  The NUV band in GALEX spans roughly 1000 \AA with an effective wavelength of 2271 \AA.  Figure \ref{galex} plots the resulting flux in $\mu$Jy versus time and orbital phase.

The observation at BJD 2456155.9617973 represents a 6$\sigma$ increase in the NUV flux from KOI-256 and may be a flare event.  The slow increase in NUV flux from BJD 2456165 to 2456180 may be due to general increase in chromospheric activity.  With the release of quarter 14 data from the {\it Kepler} Mission, we can correlate the NUV and visible light curves.  One measurement lies within the occultation and one within the transit, however neither is significantly different than the out-of-eclipse observations.  Given the strong H$\alpha$ emission from the M dwarf, and the low white dwarf temperature we measure in Section \ref{teff}, we ascribe the NUV flux to chromospheric activity in the M dwarf and not the white dwarf component.

\section{System Parameters}\label{Parameters}

In this section, we combine the measurements to empirically determine the M dwarf and white dwarf parameters, which are listed in Table \ref{parameters_table}.

\subsection{Masses, Radii and Orbital Parameters}\label{masses}

To determine the stellar masses, radii and system orbital parameters, we combine six measurements: the orbital period ($P$), the semi-amplitude radial velocity of the M dwarf ($K_{\rm dM}$), the occultation duration ($t_{\rm dur}$), the ingress/egress duration ($t_{\rm ing}$), the effective radius ratio parameter fitted to the transit light curve ($R_1/R_2$) and the rotational broadening of the M dwarf (\vsini)--with six unknowns--the M dwarf mass ($M_{\rm dM}$), the M dwarf radius ($R_{\rm dM}$), the white dwarf mass ($M_{\rm WD}$), the white dwarf radius ($R_{\rm WD}$), the orbital inclination ($i$) and the system semi-major axis ($a$)--in six equations.  The first equation is Equation \ref{lensing} in Section \ref{transit}, relating the effective radius ratio of the transit light curve to the stellar radii, white dwarf mass, and semi-major axis.  The second equation is Kepler's Third Law, which relates the measured orbital period to the masses and semi-major axis:

\begin{equation}
P =  2 \pi \sqrt{a^3 \over G \, ( M_{\rm dM} + M_{\rm WD})}
\end{equation}

\noindent The final four equations are described below.

Assuming spin-orbit synchronization, as evidenced by the out-of-eclipse variations in the long-cadence light curve (see Section \ref{long_cadence} and Figure \ref{long_cadence}), the rotational period and spin inclination of the M dwarf match the orbital period and orbital inclination of the binary.  In this case, the measured \vsini of the M dwarf relates to the orbital period, orbital inclination and M dwarf radius by the following equation:

\begin{equation}
V \sin i = { 2 \pi \, R_{\rm dM} \, \sin i \over P}
\end{equation}

The measured semi-amplitude radial velocity of the M dwarf relates to the stellar masses and orbital period by the mass function.  Assuming zero eccentricity:

\begin{equation}
K_{\rm dM} = \Big({ 2 \pi \, G \over P}\Big)^{1 \over 3} {M_{\rm WD} \, \sin i \over (M_{\rm WD} + M_{\rm dM})^{2 \over 3} }
\end{equation}

The measured occultation duration (mid-ingress to mid-egress) relates to the system parameters by the following equation:

\begin{equation}
t_{\rm dur} = { 2 \, R_{\rm dM} \, \sin i \, \sqrt{ 1 - b^2} \over K_{\rm dM} \, ( 1 + M_{\rm dM} / M_{\rm WD} )  }
\end{equation}

\noindent where $b=a \, \cos i / R_{\rm dM}$ is the impact parameter of the white dwarf relative to the center of the projected M dwarf on the sky, in units of the M dwarf radius.  And finally, the measured ingress/egress time ($t_{\rm ing}$) follows the following equation, treating the limb of the projected image of the M dwarf as a straight edge, which the white dwarf passes across at non-normal incidence:

\begin{equation}
t_{\rm ing} = {2 \, R_{\rm dM} \, \sin i \over  K_{\rm dM} \, ( 1 + M_{\rm dM} / M_{\rm WD} ) \, \sqrt{ 1 - b^2 } }
\end{equation}

We solve for the six unknowns using Broyden's Method, a secant method for solving nonlinear simultaneous equations \citep{Broyden1965}.  To estimate the uncertainties in the solved parameters due to errors in the measurements, we take a Monte Carlo approach.  We created 1000 copies of the measured values and added normally-distributed noise corresponding to the measurement errors to each copy.  For each noise-added copy of the measured parameters, we then solved for the unknowns, and take the standard deviations in the resulting 1000 iterations as the 1$\sigma$ uncertainties.  We report the resulting stellar masses, radii, orbital parameters and corresponding uncertainties in Table \ref{parameters_table}.

\subsection{White Dwarf Effective Temperature}\label{teff}

\begin{figure*}[]
\begin{center}
\includegraphics[width=6.5in]{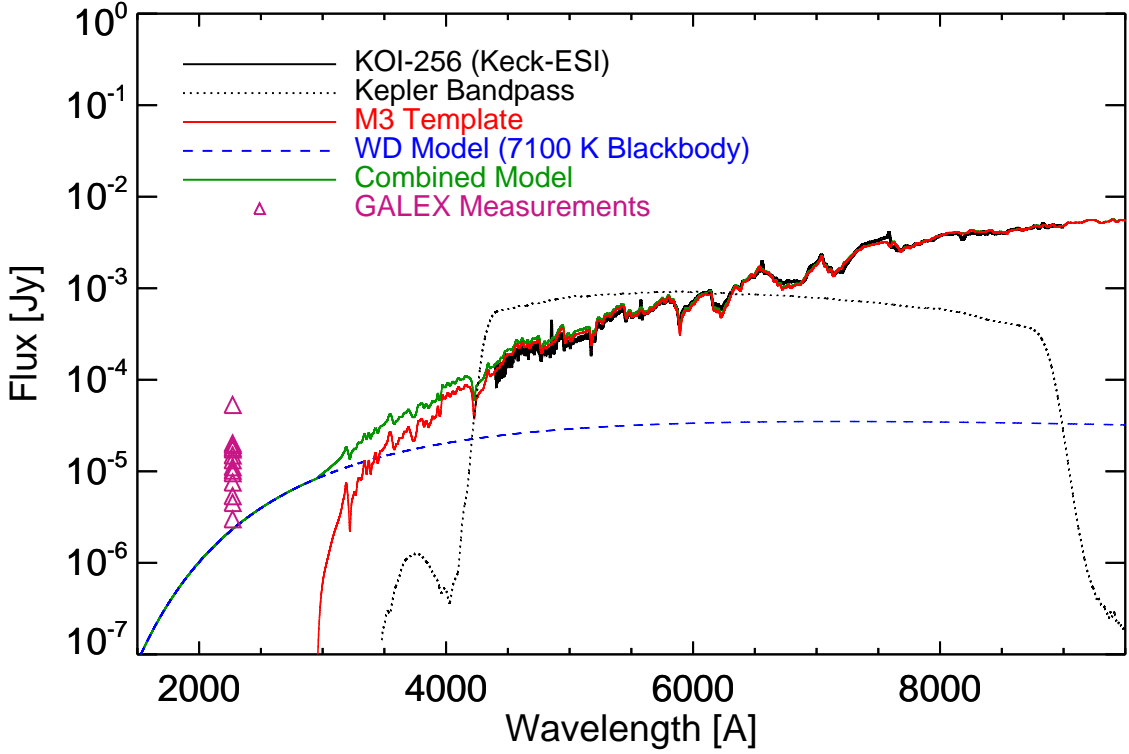}
\caption{Observed and modeled spectra of KOI-256.  The Keck-ESI spectrum ({\it solid black line}) is especially noisy from 4000 to 5000 \AA.  The model spectra are computed using the surface brightness of an M3 dwarf, calibrated using the angular diameter measurement of Gl 581 \citep[][]{vonbraun2011} and a Plank function with a temperature of 7100 $K$ for the white dwarf.  We then scaled the surface brightness to the expected flux using the stellar radii measurements described in Section \ref{masses} and KOI-256's r-band magnitude in the \ik Input Catalog \citep[KIC,][]{Batalha2010}.  The GALEX flux measurements are inconsistent with the photosphere of either the M dwarf or the white dwarf, and are likely enhanced by the chromosphere of the M dwarf.\label{esi}}
\end{center}
\end{figure*}

The white dwarf is too cool to detect in the blue wavelengths of the Keck ESI spectrum (Figure \ref{esi}), and the GALEX measurements are likely contaminated by UV emission from the active M dwarf chromosphere.  We choose to measure the white dwarf temperature using the occultation depth in the \ik band.  The fractional occultation depth ($\delta_C$) equals the ratio of the flux from the white dwarf to the total flux from the system within the \ik bandpass:

\begin{equation}\label{teff_equation}
\small
\delta_C = { {\bigintsss T_{\rm Kep,\lambda} \, S_{\rm WD, \lambda} \, R_{\rm WD}^2 \, \lambda \, d\lambda } \over { {\bigintsss ( T_{\rm Kep,\lambda} \, S_{\rm WD, \lambda} \, R_{\rm WD}^2  + T_{\rm Kep,\lambda} \, S_{\rm dM, \lambda} \, R_{\rm dM}^2 ) \, \lambda \, d\lambda }  } }
\end{equation}

Where $T_{\rm Kep,\lambda}$ is the transmission of the \ik photometer versus wavelength, $S_{\rm WD, \lambda}$ is the surface brightness of the white dwarf per unit wavelength, and $S_{\rm dM, \lambda}$ is the surface brightness of the M dwarf per unit wavelength.  The surface brightnesses of the two bodies depend on their respective temperatures and opacities over a wide range of wavelengths.  With a model for the surface brightness of the M dwarf that is consistent with the measured ESI spectrum, and a temperature-dependent model for the surface brightness of the white dwarf, we can vary white dwarf temperature to match the occultation depth in the \ik bandpass.

For the M dwarf, synthetic surface brightnesses are a matter of significant debate.  Model atmospheres have historically been unable to match spectroscopic observations due to the complicated molecular opacities in their atmospheres.  The gold standard has been the PHOENIX model atmosphere code \citep[e.g.][]{Hauschildt1999}, which reproduce infrared spectra of M dwarfs relatively well \citep[][]{Rojas2010, Rojas2012}.  However, at optical wavelengths the spectra show discrepancies from observations \citep[e.g.][]{Allard2012}.

Instead of using a model surface brightness for the M dwarf, we chose to use template spectra from \citet{Pickles1998} that have been flux-calibrated to match M dwarfs with accurate angular diameter measurements using optical long-baseline interferometry.  \citet{Boyajian2012} measured angular diameters for GJ 411 (M3V) and GJ 699 (M4V) and \citet{vonbraun2011} measured the angular diameter of Gl 581 (M3V), both using the CHARA Array on Mt. Wilson and correcting for effects from limb-darkening.  In their respective papers, they flux-calibrated M2, M3 and M4 template spectra measured by \citet{Pickles1998} to match the visible and infrared photometry of these stars.  By dividing the flux calibrated template spectra by the measured angular diameters measured for the star, we create semi-empirical surface brightness templates for the M dwarf in KOI-256.  We use these three stars, as they have similar spectral type to the M dwarf in KOI-256.

\begin{figure}[]
\begin{center}
\includegraphics[width=3.3in]{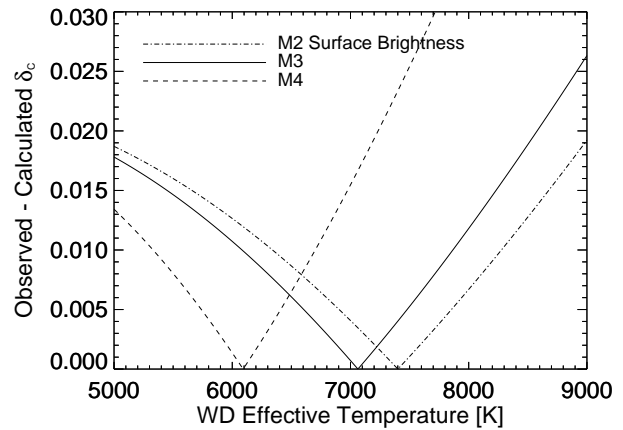}
\caption{Absolute value of the observed minus calculated \ik fractional occultation depth versus assumed white dwarf effective temperature.  We use the Planck function to model the white dwarf surface brightness, and the flux calibrated spectra and angular diameters of Gl 411 (M2, {\it dot-dashed line}), Gl 581 (M3, {\it solid line}) and Gl 699 (M4, {\it dashed} line) from \citet{vonbraun2011} and \citet{Boyajian2012} to model the M dwarf surface brightness.  We solve for the white dwarf effective temperature numerically, using the observed minus calculated \ik fractional occultation depth as a goodness-of-fit parameter.\label{teff_vs_depth}}
\end{center}
\end{figure}

For the white dwarf, we used the Planck function to model the surface brightness.  To determine the white dwarf temperature, we computed the expected occultation depth as a function of the Planck function temperature in Equation \ref{teff_equation}.  The absolute value of the difference between the calculated and measured occultation depth was used as a goodness-of-fit statistic.  The best fitting temperatures were found to be 6100 $K$, 7100 $K$ and 7400 $K$ using the M2, M3 and M4 surface brightness templates for the M dwarfs respectively (see Figure \ref{teff_vs_depth}).  We chose the value 7100 $K$ as the best estimate because the spectral type of the M dwarf in KOI-256 is closest to M3, and take the standard deviation in the three estimates as the uncertainty (700 $K$).  Figure \ref{esi} plots the measured spectrum of KOI-256 with the model spectra, \ik bandpass and GALEX flux measurements.

\begin{table*}[]
\begin{center}
\caption{Measured and Calculated Parameters for KOI-256\label{parameters_table}} 
\begin{tabular}{l c r} 
\hline\hline                        
Parameter & Value & Notes and Relevant Section\\ [0.5ex] 
\hline       
\multicolumn{3}{l}{\bf Spectroscopic Parameters} \\
M dwarf semi-amplitude RV ($K_{\rm dM}$) &    106.5 $\pm$      1.8 km $\rm s^{-1}$ & \ref{rvs}\\
Orbital eccentricity & $< 0.05$  & Assumed to be 0.0, \ref{rvs}\\
M dwarf \vsini & $19.79  \pm 0.52$ km $\rm s^{-1}$ &  \ref{hires}\\

\\
\multicolumn{3}{l}{\bf Occultation Parameters} \\
Orbital period [days] & 1.3786548 $\pm$ 0.000010 &  \ref{period}\\
Mid-occultation ephemeris (BJD) &   2455373.635498 $\pm$         0.000036 &  \ref{period}\\
Occultation duration ($t_{\rm dur}$) &  60.051 $\pm$    0.037 minutes &  \ref{occultation}\\
Ingress/egress duration ($t_{\rm ing}$)&  1.604 $\pm$    0.067 minutes &  \ref{occultation}\\
{\it Kepler} fractional depth ($\delta_C$) & 0.0238959 $\pm$ 0.0000047 &  \ref{occultation}\\
$H$-band fractional depth & $<$ 0.00275 &  3$\sigma$ upper limit, \ref{wirc}\\
\\
\multicolumn{3}{l}{\bf Transit Parameters} \\
Effective radius ratio $(R_1 / R_2)$ & 0.0209 $\pm$ 0.0011 & \ref{transit}\\
\\
\multicolumn{3}{l}{\bf M dwarf Parameters} \\
Mass ($M_{\rm dM}$)  &    0.51 $\pm$    0.15 $M_\Sun$ &  \ref{masses}\\
Radius ($R_{\rm dM}$)  &    0.540 $\pm$    0.014  $R_\Sun$ &  \ref{masses}\\
Effective temperature  & 3450 $\pm$ 50 $K$& from \citet{Muirhead2012b}\\
Metallicity [M/H] & $+0.31$ $\pm$ 0.10 & from \citet{Muirhead2012b}\\
\ik magnitude ($K_p$) & 15.40 & KIC and occultation depth\\
$K_p - H$ & 3.37 & KIC\\
\\
\multicolumn{3}{l}{\bf White Dwarf Parameters} \\
Mass ($M_{\rm WD}$) &    0.592 $\pm$    0.084 $M_\Sun$ &  \ref{masses}\\
Radius ($R_{\rm WD}$)  &    0.01345 $\pm$    0.00091 $R_\Sun$ &  \ref{masses}\\
Einstein radius ($R_{\rm Ein}$)  &    0.00473 $\pm$  0.00055 $R_\Sun$ &  \ref{masses}\\
WD \teff  & 7100 $\pm$ 800 $K$ &  \ref{teff}\\
\ik magnitude ($K_p$) & 19.45 & KIC and occultation depth\\
$K_p - H$ & $<$ 1.05 & $H$-band occultation upper limit\\
\\
\multicolumn{3}{l}{\bf Orbital Parameters} \\
Combined semi-major axis ($a$)  &  0.0250 $\pm$  0.0018 AU &  \ref{masses}\\
Orbital and M-Dwarf-Spin Inclination  &    $89.^{\circ}01$ $\pm$     0.65 &  \ref{masses}\\
$a / R_{\rm dM}$ & 9.97 $\pm$ 0.64 & \ref{masses}\\
Impact parameter ($b$) & 0.17 $\pm$ 0.10 & \ref{masses}\\

\\
\multicolumn{3}{l}{\bf Assumptions} \\
\multicolumn{2}{l}{The M dwarf is synchronously rotating with orbit.} & Evidence in \ref{avi} \\
\multicolumn{2}{l}{The M dwarf spin axis is aligned with binary orbital axis.} & Evidence in \ref{avi}\\
\multicolumn{3}{l}{The M dwarf obeys \citet{Claret2011} limb-darkening parameters.}\\
\hline 
\end{tabular}
\end{center}
\end{table*}

\section{Discussion}\label{Discussion}

KOI-256 is one of several eclipsing systems containing a white dwarf discovered in the \ik field. Four of them, KOI-74 and KOI-81 \citep{rowe2010, vankerkwijk2010}, KIC 10657664  \citep{Carter2011c}, and KOI-1224 \citep{Breton2012} are composed of a low-mass, hot white dwarf, with $M_{\rm WD} \sim 0.2 M_\Sun$ and \teff $>$ 10,000 $K$, and an early-type main sequence star. Two other systems have a hot WD whose mass is close to the canonical mass of 0.6 ${\rm M_{\sun}}$. Those are KPD 1946+4340 \citep{Bloemen2011}, where the WD is orbiting a sub-dwarf B star, and KIC 10544976 \citep{Almenara2012}, with an M dwarf primary. The latter two have an orbital period shorter than 10 hrs, compared to the 33 hr period of KOI-256. Therefore, KOI-256 is the first cold WD discovered in the Kepler field orbiting a low-mass star at a relatively long period.  KOI-256 is also a model example of gravitational lensing effects during transit.  With empirically measured masses and radii, KOI-256 serves as a benchmark object for studying the effects of PCEB evolution on M dwarf abundance and activity as a function of mass and radius.

\subsection{The Evolution of the System}\label{evolution}

With its short orbital period and white dwarf secondary, KOI-256 is unquestionably a PCEB and will eventually evolve into a cataclysmic variable. However, KOI-256 has a relatively long orbital period compared to most known \textquotedblleft pre-CVs" \citep{Schreiber2003}, and it contains a relatively cool white dwarf. These characteristics indicate KOI-256 is relatively old.  The time elapsed since the end of the common envelope phase can be estimated from the white dwarf cooling time. Using the cooling tracks of Renedo et al. (2010), we estimate a cooling time $t_{\rm cool} \approx 2$ Gyr for the white dwarf in KOI-256. We also use the initial to final mass relationships of \citet{Renedo2010} to estimate that the white dwarf in KOI-256 originated from a progenitor of initial mass $1.5 M_\odot \lesssim M_i \lesssim 3M_\odot$, although the initial to final mass relationship is less certain for white dwarfs in PCEB binaries \citep{Rebassa-Mansergas2011}.

The orbits of pre-CVs decay primarily due to angular momentum loss from the dwarf star's magnetized wind. We use the angular momentum loss prescription of Sills et al. (2000) and Andronov (2003), in which the system's angular momentum, $J$, changes as
\begin{equation}
\label{jdot}
\frac{d J}{dt} = - K \bigg(\frac{M_\odot}{M}\bigg)^{1/2} \bigg(\frac{R}{R_\odot}\bigg)^{1/2} \omega_c^2 \ \omega
\end{equation}
where $K=2.7 \times 10^{47} {\rm g} \ {\rm cm}^2 \ {\rm s}$ is a calibrated constant, $\omega$ is the angular spin frequency of the dwarf star, and $\omega_c$ is the angular spin frequency at which the angular momentum loss saturates (we use $\omega_c = 7 \omega_\odot$). Assuming efficient spin-orbit coupling and ignoring the small angular momentum contained in the stars, the initial orbital frequency of the system was
\begin{equation}
\label{omi}
\Omega_i \simeq \bigg[ \frac{4 A t_{\rm cool}}{\mu (G M_t)^{2/3}} + \Omega_0^{-4/3} \bigg]^{-3/4},
\end{equation}
where $\mu$ is the reduced mass of the system, $M_t$ is its total mass, $A= K [R M_\odot/(M R_\odot)]^{1/2} \omega_c^2$, and $\Omega_0$ is the current orbital angular frequency. We use equation (\ref{omi}) to estimate that KOI-256 emerged from the common envelope phase with an orbital period of $P_i \approx 1.7$ days. The elapsed time until Roche lobe overflow is
\begin{equation}
\label{trl}
t_{RL} \simeq \frac{\mu (G M_t)^{2/3}}{4 A} \Big(\Omega_0^{-4/3} - \Omega_{RL}^{-4/3}\Big),
\end{equation}
where $\Omega_{RL}$ is the angular orbital frequency at Roche lobe overflow. We estimate mass transfer will commence at an orbital period of $P_{RL} \simeq 280$ minutes in KOI-256, and that this will occur after $t_{RL} \approx 5.3$ Gyr of orbital decay. After the initiation of mass transfer, KOI-256 will likely evolve as a fairly typical CV system. 

We can also calculate whether the orbital decay in KOI-256 is likely to be observed via eclipse timing. Over an observational baseline $T$, the eclipse times change by an amount
\begin{equation}
\label{deltat}
\Delta t = \frac{\dot{\Omega} T^2}{2 \Omega}.
\end{equation}
In KOI-256, we estimate the eclipse times will only change by $\Delta t \approx 1$ s over the next twenty years, so it is unlikely that the orbital decay will be detected in the near future.

\subsection{On the High Metallicity of the M Dwarf}

\citet{Muirhead2012b} measured an overall metallicity of [M/H] = +0.31 $\pm$ 0.10 using the $K$-band near-infrared spectroscopic indices of \citet{Rojas2010, Rojas2012}.  The indices are calibrated on widely separated main-sequence binary stars, and the calibration may not be applicable to the M dwarf in KOI-256 due effects from common-envelope evolution.  Nevertheless, the high metallicity compared to the other M dwarf KOIs analyzed in that paper (mean [M/H] = -0.11, RMS [M/H] = 0.15) is interesting considering that nearby M dwarfs with white dwarf companions also show high metallicities using the $K$-band calibration.  \citet{Rojas2012} found that G 203-47 and GL 169.1 A, both nearby M dwarfs with white dwarf companions, had  metallicities of [M/H] = 0.24 $\pm$ 0.12 and 0.26 $\pm$ 0.12, respectively.  They speculate that the high metallicities could be due to enhancement by nucleosynthesis products from the white-dwarf progenitor during the common-envelope phase.  If true, the high metallicity of M dwarfs in PCEBs maybe a concern for metallicity-biased exoplanet surveys of M dwarfs since their high metallicity is not primordial and therefore does not necessarily indicate an increased likelihood of planet occurrence \citep[e.g.][]{Apps2010}.

\subsection{Future Observations}\label{future}

Future investigation of KOI-256 will enable ever more precise measurements the parameters of the system, including the mass, radius, and temperature of both objects, which are especially interesting in the case of the white dwarf. These investigations will benefit from follow-up observations beyond future \ik data, including UV spectroscopy and multi-band high-speed occultation photometry. The latter will constrain the radii and temperature ratio, and may be able to detect the asymmetry in the occultation ingress and egress light curves due to the Photometric Rossiter-McLaughlin effect \citep{Shporer2012,Groot2012}. Modeling this effect will enable measuring the white dwarf obliquity and rotational velocity.

\acknowledgements

We would like to thank Eric Agol, Jean-Michel D\'{e}sert, Dong Lai, Dylan Morgan, Tony Piro and Andrew West for their insightful communications about this system.  We would like to thank the anonymous referee for the thoughtful and constructive comments.  We would like to thank the staff at Palomar Observatory for providing support during our many observation runs, including Bruce Baker, Mike Doyle, Jamey Eriksen, Carolyn Heffner, John Henning, Steven Kunsman, Dan McKenna, Jean Mueller, Kajsa Peffer, Kevin Rykoski, and Greg van Idsinga.  J. Becker would like to thank Mr. and Mrs. Kenneth Adelman for providing funding for her 2012 Alain Porter Memorial SURF Fellowship.  A portion of this work was supported by the National Science Foundation under grant No. AST-1203023.

The Robo-AO system is supported by collaborating partner institutions, the California Institute of Technology and the Inter-University Centre for Astronomy and Astrophysics, by the National Science Foundation under grant Nos. AST-0906060 and AST-0960343, by a grant from the Mt. Cuba Astronomical Foundation and by a gift from Samuel Oschin.  This paper includes data collected by the \ik mission. Funding for the \ik mission is provided by the NASA Science Mission directorate.  Some of the data presented in this paper were obtained from the Mikulski Archive for Space Telescopes (MAST). STScI is operated by the Association of Universities for Research in Astronomy, Inc., under NASA contract NAS5-26555. Support for MAST for non-HST data is provided by the NASA Office of Space Science via grant NNX09AF08G and by other grants and contracts.

Some of the data presented herein were obtained at the W.M. Keck Observatory, which is operated as a scientific partnership among the California Institute of Technology, the University of California and the National Aeronautics and Space Administration. The Observatory was made possible by the generous financial support of the W.M. Keck Foundation.  The authors wish to recognize and acknowledge the very significant cultural role and reverence that the summit of Mauna Kea has within the indigenous Hawaiian community.  

{\it Facilities:} \facility{GALEX}, \facility{Keck:I (HIRES)}, \facility{Keck:II (ESI)}, \facility{Kepler}, \facility{PO:1.5m (Robo-AO)}, \facility{PO:Hale (TripleSpec, WIRC)}

\clearpage
\bibliographystyle{apj}
\bibliography{koi256}

\end{document}